# Effect of hydrostatic pressures on the superconductivity of new $BiS_2$ based $REO_{0.5}F_{0.5}BiS_2$ (RE-La, Pr and Nd) superconductors

Rajveer Jha, H. Kishan and V. P. S. Awana[*]

CSIR-National Physical Laboratory, Dr. K.S. Krishnan Marg, New Delhi-110012, India

We study the impact of hydrostatic pressure on superconductivity of new $BiS_2$ based layered $REO_{0.5}F_{0.5}BiS_2$ (RE-La, Pr, and Nd) compounds through the measurements of dc electrical resistivity. The $REO_{0.5}F_{0.5}BiS_2$ (RE-La, Pr and Nd) compounds synthesized by solid state reaction route via vacuum encapsulation are crystallized in the tetragonal P4/*nmm* space group. At ambient pressure the superconducting transition onset temperatures ($T_c^{onset}$) are 2.7K, 3.5K and 4.5K which are enhanced substantially under external hydrostatic pressure to 10.5K, 7.8K and 7.5K for $LaO_{0.5}F_{0.5}BiS_2$, $PrO_{0.5}F_{0.5}BiS_2$ and $NdO_{0.5}F_{0.5}BiS_2$ respectively at 1.68GPa. The normal state electrical resistivity decreases with applied pressure for $REO_{0.5}F_{0.5}BiS_2$ (RE-La, Pr and Nd). The electrical resistivity under magnetic field and applied pressure has been measured to estimate upper critical field $H_{c2}(0)$, the values of which are 15.9Tesla, 8.8Tesla and 8.2Tesla for $LaO_{0.5}F_{0.5}BiS_2$, $PrO_{0.5}F_{0.5}BiS_2$ and $NdO_{0.5}F_{0.5}BiS_2$ compounds. Substantial enhancement of superconductivity under moderate pressures in studied new $BiS_2$ based superconductors call for the attention of condensed matter physics community.

**Introduction:**

Since the discovery of superconductivity in layered $BiS_2$ based superconductors $Bi_4O_4S_3$ with $T_c$ =4.5K [1, 2] and $REO_{1-x}F_xBiS_2$ (RE=La, Ce, Pr and Nd) with $T_c$ =3K -10K [3-10], lot of studies have been done by both experimentalists and theoreticians. These compounds consist of two dimensional $BiS_2$ layers with different charge reservoir blocks in structure as similar to FeAs in Fe- based and $CuO_2$ in high $T_c$ cuprate superconductors [11, 12]. Basically, parent phase of $BiS_2$ based superconductors exhibit the Mott-insulating or semiconducting behavior. According to some theoretical studies, the fundamental electronic properties are attributed to the $BiS_2$ layers [1-10]. The band structure can be described as the valance band being consisting of $S_{3p}$ and $O_{2p}$ orbitals and the conduction band is mainly attributed to $Bi6p_{xy}$ and $S_{3p}$ [13-15]. The electron doping is achieved through the $O^{-2}$ site $F^{-1}$ substitution. The weak correlation effect in p orbitals has been suggested by some band structure calculation studies. Also it is suggested that possibly the electron phonon coupling plays an important role in superconducting pairing in these systems [13–17].

Superconductivity has been observed in the layered $BiS_2$ based compounds through the onsite chemical substitutions. For example, the $O^{2-}$ site $F^{-1}$ doped $REOBiS_2$ compounds show superconductivity in range of 2.5 to 5K, and $Sr^{2+}$ site $La^{+3}$ doped $SrFBiS_2$ exhibit superconductivity near 2.5K [18-20]. The superconductivity has also been observed in $LaOBiS_2$ by chemical substitution of the tetravalent ions viz. $Th^{+4}$, $Hf^{+4}$, $Zr^{+4}$ and $Ti^{+4}$ for trivalent $La^{+3}$ [21]. Basically, it is the doping of mobile carriers in conducting $BiS_2$ layer via $O^{2-}/F^{-1}$, $Sr^{2+}/La^{3+}$ or $La^{3+}/Th^{4+}$, $Hf^{4+}$, $Zr^{4+}$ substitution in charge reservoir REO layer. The doping mechanism seems pretty same to that as in case of high $T_c$ cuprate and Fe based pnictide superconductors [11,12]. The role of Rare earth elements is also crucial in superconductivity of $BiS_2$ based compounds similar to that as for Fe based high $T_c$ superconductors. Some experimental studies showed that superconductivity effectively increases as light rare earth element is replaced by the heavier one [3-10]. The pressure dependent studies have also been done, which resulted in modifying the superconducting properties of these $BiS_2$ based superconductors [22-25]. The externally applied pressure may effectively change in lattice constants through bond lengths and angles and that could affect the electronic and magnetic correlations of a superconductor [24]. In addition, the structural phase transition from tetragonal to monoclinic has also been reported on the basis of analysis of the x-ray diffraction patterns under high pressure for the $LaO_{0.5}F_{0.5}BiS_2$ superconductor [16]. More recently perfect tetragonal structure has been reported for the high pressure synthesized of $CeO_{0.3}F_{0.7}BiS_2$ compound [27]. In any case, the pressure effect studies on the $BiS_2$ superconductors need to be explored further by different research groups, because yet only scant reports do exist [22-25].

Here we present the effect of pressure on the superconductivity of the $REO_{0.5}F_{0.5}BiS_2$ (La, Pr & Nd). At zero pressure the normal state resistivity of all the compounds show semiconducting behaviour. As the pressure increases the normal state resistivity is suppressed for $LaO_{0.5}F_{0.5}BiS_2$ and $PrO_{0.5}F_{0.5}BiS_2$ compounds and slightly increased initially for $NdO_{0.5}F_{0.5}BiS_2$ compound. Further their superconducting properties are enhanced dramatically under moderate hydrostatic pressures and the impact of the same is most in case of $LaO_{0.5}F_{0.5}BiS_2$. The electrical resistivity under magnetic field and applied pressure gave the estimated $H_{c2}(0)$ values of 15.9Tesla, 8.8Tesla and 8.2Tesla for $LaO_{0.5}F_{0.5}BiS_2$, $PrO_{0.5}F_{0.5}BiS_2$ and $NdO_{0.5}F_{0.5}BiS_2$ compounds. Interestingly the estimated $H_{c2}(0)$ values are close to Pauli paramagnetic limit of $1.84T_c$, indicating the robustness of superconductivity in $BiS_2$ based new superconducting systems.

**Experimental Details:**

The bulk polycrystalline $REO_{0.5}F_{0.5}BiS_2$ (RE=La, Pr, & Nd) samples have been synthesized by solid state reaction route via vacuum encapsulation. High purity chemicals La, $La_2O_3$, $LaF_3$, Bi and S are used for $LaO_{0.5}F_{0.5}BiS_2$; Pr, $Pr_6O_{11}$, $PrF_3$, Bi, and S are used for $PrO_{0.5}F_{0.5}BiS_2$ and Nd, $Nd_2O_3$, $NdF_3$, Bi and S are used for $NdO_{0.5}F_{0.5}BiS_2$ compound. The chemicals are weighed in stoichiometric ratio and ground in glove box filled with high purity Argon. The mixed powder was subsequently palletized and vacuum-sealed ($10^{-4}$ mbar) in a quartz tube. The box furnace have been used to sinter the samples at $650^0C$ for 12h with the typical heating rate of $2^oC$/minute and subsequently cooled down slowly to room temperature. This process was repeated twice. X-ray diffraction (*XRD*) was performed at room temperature in the scattering angular (*2θ*) range of $10^o$-$80^o$ in equal *2θ* step of $0.02^o$ using *Rigaku Diffractometer* with *Cu $K_α$* ($λ = 1.54Å$). Rietveld analysis was performed using the standard *FullProf* program.

The pressure dependent resistivity measurements were performed on Physical Property Measurements System (*PPMS*-14T, *Quantum Design*) by using HPC-33 Piston type pressure cell with Quantum design DC resistivity Option. Hydrostatic pressures were generated by a BeCu/NiCrAl clamped piston-cylinder cell. The sample was immersed in a fluid pressure transmitting medium of Fluorinert (FC70:FC77=1:1) in a Teflon cell. Annealed Pt leads were affixed to gold-sputtered contact surfaces on each sample with silver epoxy in a standard four-wire configuration. The pressure at low temperature was calibrated from the superconducting transition temperature of Pb.

**Results and discussion:**

Figure 1(a) shows the room temperature observed and Reitveld fitted XRD pattern of $LaO_{0.5}F_{0.5}BiS_2$, $PrO_{0.5}F_{0.5}BiS_2$ and $NdO_{0.5}F_{0.5}BiS_2$ compounds. All the compounds are crystallized in tetragonal structure with space group *P4/nmm*. Small impurity peaks, close to the background of $Bi_2S_3$ and Bi are also observed in all the three compounds. All the studied samples are phase pure within the XRD limit. Reitveld fitted lattice parameters are $a=4.06(2)$Å, $c=13.38(1)$Å for $LaO_{0.5}F_{0.5}BiS_2$; $a=4.0(2)$Å, $c=13.42(1)$Å for $PrO_{0.5}F_{0.5}BiS_2$ and $a=3.99(3)$Å, $c=13.38(1)$Å for $NdO_{0.5}F_{0.5}BiS_2$ compound. The volume of the unit cell obtained through the Reitveld fitting analysis of each compound is $221.71(1)$Å$^3$ for $LaO_{0.5}F_{0.5}BiS_2$, $216.33(1)$Å$^3$ for $PrO_{0.5}F_{0.5}BiS_2$ and $214.60(1)$Å$^3$ for $NdO_{0.5}F_{0.5}BiS_2$. The volume of unit cell is decreased from $221.71(1)$Å$^3$ to $214.60(1)$Å$^3$ with the decrease in atomic radii of the RE element. Figure 1(b) shows the schematic unit cell of the $REO_{0.5}F_{0.5}BiS_2$ compounds, which has been drown from Reitveld refined parameters. The details are discussed elsewhere [4,6,7].

Figure 2(a) shows the temperature dependent electrical resistivity from 300K down to 2K at various applied pressures for the $LaO_{0.5}F_{0.5}BiS_2$ compound. At ambient pressure normal state resistivity ρ(T) exhibits semiconducting behaviour, which is strongly suppressed with increasing pressure from 0-0.97GPa and later nearly unaltered for the higher applied pressures of above 1GPa. Figure 2(b) depicts the zoom part of the ρ(T) from 14K down to 2K near the superconducting transition regime at various pressures. It can be clearly seen that the superconducting temperature $T_c(ρ=0)$ is 2K at 0GPa pressure, which increases to 2.5K at applied pressure of 0.35GPa. Interestingly for the applied pressure of 0.55GPa, the $T_c(ρ=0)$ and $T_c^{onset}$ increase to 5K and 9.8K respectively with broadening of the superconducting transition. For the applied pressure of 0.97GPa, though the $T_c^{onset}$ is increased only marginally to 10.5K, the $T_c(ρ=0)$ is increased sharply to 9.5K, with relatively sharper superconducting transition. For the further higher applied pressures of 1.35GPa and 1.68GPa, though the $T_c^{onset}$ and $T_c(ρ=0)$ are nearly unchanged, but the normal state resistivity is still being suppressed progressively. The obtained highest $T_c^{onset}$ is 10.5K for the applied pressure of 0.97GPa. The pressure coefficient $dT_c^{onset}/dP$ for $LaO_{0.5}F_{0.5}BiS_2$ is estimated to be around 8.25K/GPa at 0.97GPa.

Plots of the temperature dependence of the electrical resistivity ρ(T) below 300K for $PrO_{0.5}F_{0.5}BiS_2$ at various pressures of 0 to 1.97GPa are shown in Figure 3(a). The compound exhibits semiconducting behavior. The normal state resistivity is strongly suppressed with applied pressures similar to $LaO_{0.5}F_{0.5}BiS_2$ compound, but yet the compound exhibits the semiconducting behaviour, till highest studied pressure of 1.97GPa. Figure 3(b) shows the $\rho(T)$ curve in the temperature range of 10K to 2K to visualise the superconducting transition temperature of the compound at various pressures. It is clearly seen that with applied pressures the $T_c^{onset}$ increases from 3.5K to 6.5K at 1.38GPa and the $T_c(\rho=0)$ is increased slightly from around 3K to 4K. For further higher applied pressures of 1.68 and 1.97GPa, both the $T_c^{onset}$ and $T_c(\rho=0)$ are saturated at around 7.5K and 6K respectively. The estimated $dT_c^{onset}/dP$ for $PrO_{0.5}F_{0.5}BiS_2$ is ~2.38K/GPa.

Figure 4(a) represents $\rho(T)$ below 300K for $NdO_{0.5}F_{0.5}BiS_2$ at various pressures from 0 to 1.97GPa. At the ambient pressure normal state resistivity $\rho(T)$ exhibits semiconducting behaviour. Both normal state resistivity and the semiconducting behaviour are suppressed initially with applied pressure. For higher pressures though the decrease in resistivity is not monotonic and rather the same increase slightly, but the semiconducting behaviour keeps on improving. Figure 4(b) shows the expanded part of Figure 4(a) in superconducting regime from 10K to 3K, to clearly visualize the change in $T_c$ with pressure. The $NdO_{0.5}F_{0.5}BiS_2$ compound shows the $T_c(\rho=0)$ at 4.1K for 0GPa pressure and the same slightly increases up to 4.3K for the applied pressure of 0.35 and 0.55GPa. At 0.97GPa applied pressure, the $T_c(\rho=0)$ and $T_c^{onset}$ increase to 4.5K and 6K respectively, with slight broadening of the superconducting transition. For the applied pressure of 1.38GPa the superconducting transition temperature is nearly the same but the $T_c^{onset}$ is increased and the resistivity behaviour is changed. For still higher applied pressure of 1.68 and 1.97GPa the $T_c(\rho=0)$ of the sample is increased to 6K and the $T_c^{onset}$ to 7.5K. The pressure coefficient $dT_c^{onset}/dP$ for $NdO_{0.5}F_{0.5}BiS_2$ is thus nearly 1.8K/GPa.

The Pressure dependence of superconducting transition temperature $T_c(\rho=0)$ for the studied $LaO_{0.5}F_{0.5}BiS_2$ $PrO_{0.5}F_{0.5}BiS_2$ and $NdO_{0.5}F_{0.5}BiS_2$ compounds is plotted in Figure 5. For the $LaO_{0.5}F_{0.5}BiS_2$ compound the $T_c(\rho=0)$ is enhanced sharply at 0.97GPa. We mark this pressure as $P_T$. Interestingly, for the $PrO_{0.5}F_{0.5}BiS_2$ and $NdO_{0.5}F_{0.5}BiS_2$ compounds the $P_T$ is at 1.68GPa. Not only the $P_T$ is nearly half (0.97GPa) for $LaO_{0.5}F_{0.5}BiS_2$ than of 1.68GPa for $PrO_{0.5}F_{0.5}BiS_2/NdO_{0.5}F_{0.5}BiS_2$, the increase in $T_c(\rho=0)$ at $P_T$ is around four fold for former and only two fold for the later ones. In our studies within limited range of applied pressure of up to 1.97GPa, though we did not observe any deleterious effect on superconductivity, but the saturation is clearly seen below the studied pressures. Worth mentioning is the fact that though our results are qualitatively similar to that as reported recently by Wolowiec et al [23,24], a confirmation of such an interesting result (four fold increase in $T_c$ at just ~ 1GPa) by independent research group was warranted.

Figure 6 shows the log(ρ) vs 1/T plots for $LaO_{0.5}F_{0.5}BiS_2$ compound at various pressures of up to 1.68GPa. This is done to find out the change in values of semiconducting transport energy gaps with pressure, as proposed earlier [23, 24]. It is seen in Figs. 2, 3 and 4, that the normal state semiconducting behaviours gets suppressed with application of pressure in studied $REO_{0.5}F_{0.5}BiS_2$ (RE=La, Pr and Nd). The fitting of log(ρ) vs 1/T for transport data could effectively provide the information about change in semiconducting energy gap. It has been observed earlier that the resistivity data can be described in to distinct regions from the relation $\rho(T) = \rho_0 e^{\Delta/2k_BT}$ where $\rho_0$ is a constant, $\Delta$ is an energy gap and $k_B$ is Boltzmann constant [22]. Resistivity data in the whole temperature range is not possible to fit linearly in above relation, so we are using the same formulism as used by Kotegawa et al. [22]. They explain the two energy gaps as $\Delta_1$ and $\Delta_2$ in the temperature range 300 to 200K and from 20K to $T_c^{onset}$ respectively. The estimated values of the energy gaps are $\Delta_1/k_B \approx 201K$ and $\Delta_2/k_B \approx 11.98K$ for the $LaO_{0.5}F_{0.5}BiS_2$ compound at zero pressure. The obtained values are comparable to the previous report on the same compound [22] but significantly less than the ones reported by Wolowiec *et al.* for the $LaO_{0.5}F_{0.5}BiS_2$ compound [23].

Figures 7 (a) and (b) shows the behaviour of energy gaps at the different pressures from 0-1.68GPa for $LaO_{0.5}F_{0.5}BiS_2$ and from 0-1.97GPa for $PrO_{0.5}F_{0.5}BiS_2$ compounds. Worth mentioning is the fact that the non-monotonic change is semiconducting behaviour of $NdO_{0.5}F_{0.5}BiS_2$ with pressure did not allow the log(ρ) vs 1/T fitting for the same at least for higher pressure range of above 0.97GPa. Both the energy gaps $\Delta_1$ and $\Delta_2$ decrease with the applied pressure from 0-1.68GPa for LaO0.5F0.5BiS2, please see Fig. 7(a). The saturation in the decrement of energy gaps may be possible for the higher applied pressures, as reported by Wolowiec *et al.* [23]. We have also obtained the values of energy gaps $\Delta_1$ and $\Delta_2$ for the $PrO_{0.5}F_{0.5}BiS_2$ sample. At the ambient pressure the evaluated values of energy gaps are $\Delta_1/k_B \approx 108.3$ K and $\Delta_2/k_B \approx 4.7$ K for the $PrO_{0.5}F_{0.5}BiS_2$ compound. From the Figures 7(b) it can be observed that both the energy gaps decreases rapidly, indicating better conduction with increasing the pressure. For $PrO_{0.5}F_{0.5}BiS_2$ compound the first energy gap ($\Delta_1$) value is almost saturated at 1.97GPa. It may be possible that with application of pressure the bond angle and band structure changes, which positively affect the electronic and magnetic correlations. The application of pressure could also moderate the charge carrier density at the Fermi surface in these $BiS_2$ based systems [26]. Also the possibility of very strong electron correlations in superconducting $BiS_2$ layers need to be explored. Interestingly, in some theoretical studies an insulator to metal transition under pressure along with superconductivity at low temperatures has been suggested for the $LaO_{0.5}F_{0.5}BiS_2$ compound [18].

Figure 8 (a)-(c) represent the temperature dependent resistivity under applied magnetic fields of up to 5Tesla for $LaO_{0.5}F_{0.5}BiS_2$, at 1.68GPa, $PrO_{0.5}F_{0.5}BiS_2$ at 1.97GPa and $NdO_{0.5}F_{0.5}BiS_2$ at 1.97GPa pressure. For all the compounds the superconductivity decreases with increasing applied field as similar to any type II superconductor. Interestingly, the $LaO_{0.5}F_{0.5}BiS_2$ is relatively more robust against magnetic field than $PrO_{0.5}F_{0.5}BiS_2$ and $NdO_{0.5}F_{0.5}BiS_2$. Relative decrement in superconductivity with field i.e., $dT_c/dH$ is around 1.2K/Tesla from absolute $T_c$ ($\rho=0$) criteria for $LaO_{0.5}F_{0.5}BiS_2$, which is near about the same as for Fe Pnictides [11] and more than the High $T_c$ superconductors [12]. For $PrO_{0.5}F_{0.5}BiS_2$ and $NdO_{0.5}F_{0.5}BiS_2$, the $dT_c/dH$ is around 1.6K/Tesla and 1.79K/Tesla respectively. The upper critical field $H_{c2}(0)$ of all the samples has been estimated by using the conventional one-band Werthamer–Helfand–Hohenberg (*WHH*) equation for the 90% criterion of the normalized resistivity data, i.e., $H_{c2}(0) = −0.693T_c(dH_{c2}/dT)_{T=Tc}$. The estimated $H_{c2}(0)$ has been shown in the Fig. 9 and the values are 15.9Tesla, 8.8 Tesla and 8.2Tesla for $LaO_{0.5}F_{0.5}BiS_2$, $PrO_{0.5}F_{0.5}BiS_2$, and $NdO_{0.5}F_{0.5}BiS_2$. Interestingly enough the $H_{c2}(0)$ of around 16Tesla for $LaO_{0.5}F_{0.5}BiS_2$ compound having $T_c$ ($\rho=0$) of 9.5K is very close to the Pauli paramagnetic limit of $1.84T_c$.

**Conclusion:**

In conclusion, we reported the impact of hydrostatic pressure on the superconductivity of $BiS_2$ based new superconductors $LaO_{0.5}F_{0.5}BiS_2$, $PrO_{0.5}F_{0.5}BiS_2$ and $NdO_{0.5}F_{0.5}BiS_2$. Superconductivity is enhanced dramatically with improved normal state conduction for all the studied systems. These results call for unusual superconductivity in these systems along with strong electron correlations. The upper critical field is estimated from electrical resistivity under magnetic field and applied pressure data and found to be 15.9Tesla, 8.8 Tesla and 8.2Tesla for $LaO_{0.5}F_{0.5}BiS_2$, $PrO_{0.5}F_{0.5}BiS_2$, and $NdO_{0.5}F_{0.5}BiS_2$. It seems the $BiS_2$ based superconductors are quite robust against magnetic field. Our results are in confirmation with the very recent report by Wolowiec *et al.*[23,24], with an added feature on upper critical field of the under pressure samples. These results on new $BiS_2$ based superconductors are thought provoking and do call for the attention of condensed matter physics community.

**Acknowledgement:**

The authors would like to thank their Director Professor R C Budhani for his support and encouragement in the present work. R. Jha would like to thank CSIR for Junior Research Fellowship and H. Kishan for emeritus scientist fellowship. This work is supported by DAE-SRC outstanding investigator award scheme to work on search for new superconductor.

**Figure Captions**

**Figure 1:** (a) Reitveld fitted XRD patterns for $LaO_{0.5}F_{0.5}BiS_2$, $PrO_{0.5}F_{0.5}BiS_2$ and $NdO_{0.5}F_{0.5}BiS_2$ compounds at room temperature, (b) schematic unit cell of $REO_{0.5}F_{0.5}BiS_2$ (La, Pr & Nd) compounds.

**Figure 2:** (a) ρ Vs T plots for $LaO_{0.5}F_{0.5}BiS_2$ compound, at varying pressures in the temperature range 300K-2.0 K. (b) ρ Vs T plots for $LaO_{0.5}F_{0.5}BiS_2$ compound, at varying pressures in the temperature range 15K-2.0 K.

**Figure 3:** (a) ρ Vs T plots for $PrO_{0.5}F_{0.5}BiS_2$ compound, at varying pressures in the temperature range 300K-2.0 K. (b) ρ Vs T plots for $PrO_{0.5}F_{0.5}BiS_2$ compound, at varying pressures in the temperature range 10K-2.0 K.

**Figure 4:** (a) ρ Vs T plots for $NdO_{0.5}F_{0.5}BiS_2$ compound, at varying pressures in the temperature range 300K-2.0 K. (b) ρ Vs T plots for $NdO_{0.5}F_{0.5}BiS_2$ compound, at varying pressures in the temperature range 10K-3.0 K.

**Figure 5:** $T_c(\rho=0)$ versus pressure plots for $LaO_{0.5}F_{0.5}BiS_2$, $PrO_{0.5}F_{0.5}BiS_2$ and $NdO_{0.5}F_{0.5}BiS_2$ compounds.

**Figure 6:** ρ vs 1/T plots up to 1.68 GPa for $LaO_{0.5}F_{0.5}BiS_2$ compound. The solid lines are liner fit of $\rho(T) = \rho_0 e^{\Delta/2k_B T}$ equation at high and low temperatures.

**Figure 7:** Energy gaps $\Delta_1$ and $\Delta_2$ as a function of pressure for (a) $LaO_{0.5}F_{0.5}BiS_2$ and (b) $PrO_{0.5}F_{0.5}BiS_2$ compound.

**Figure 8:** ρ(T) under magnetic fields for the (a) $LaO_{0.5}F_{0.5}BiS_2$ at 1.68GPa, (b) $PrO_{0.5}F_{0.5}BiS_2$ at 1.97GPa, and (c) $NdO_{0.5}F_{0.5}BiS_2$ at1.97GPa

**Figure 9:** $H_{c2}(0)$ Vs T plots for the $LaO_{0.5}F_{0.5}BiS_2$ at 1.68GPa and $PrO_{0.5}F_{0.5}BiS_2$, $NdO_{0.5}F_{0.5}BiS_2$ at 1.97GPa hydrostatic pressures.

Figure 1

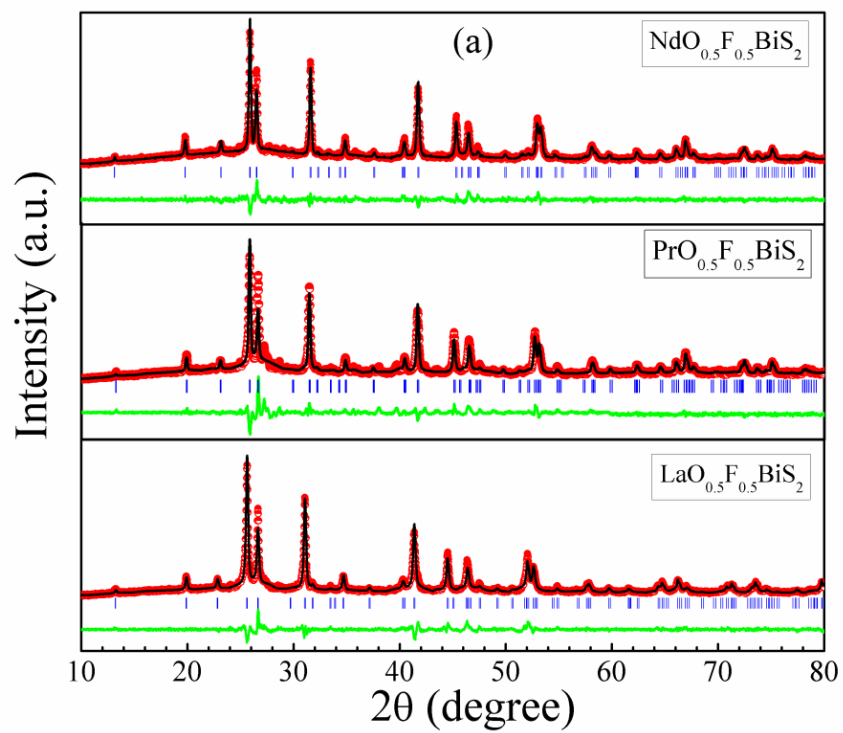

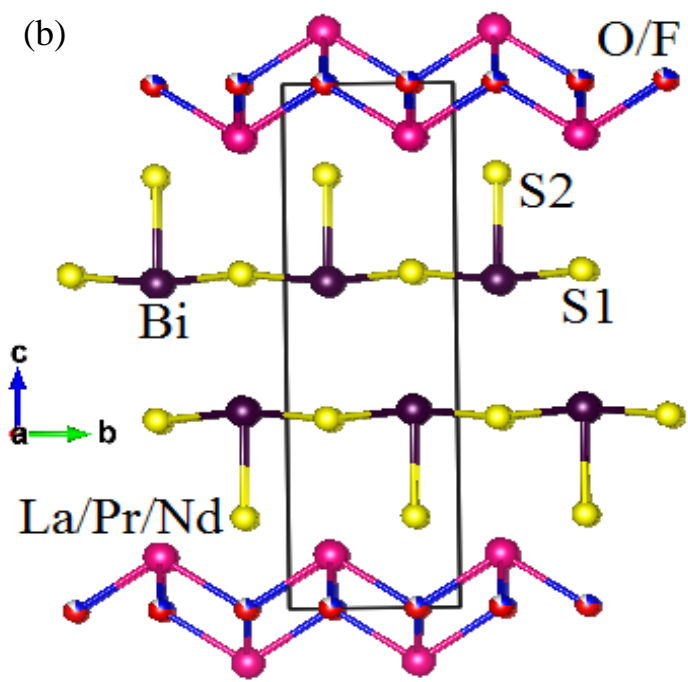

Figure 2

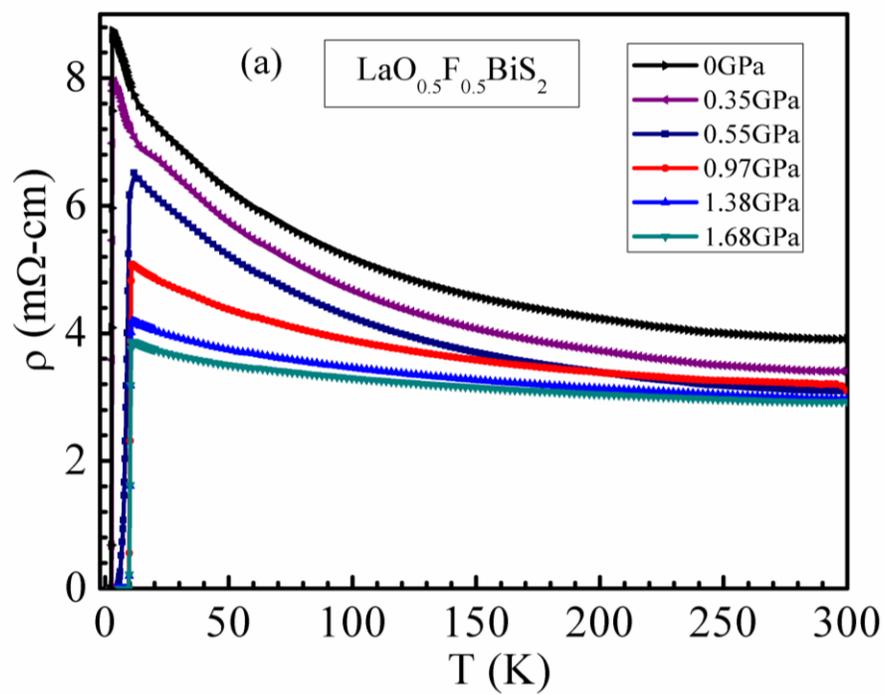

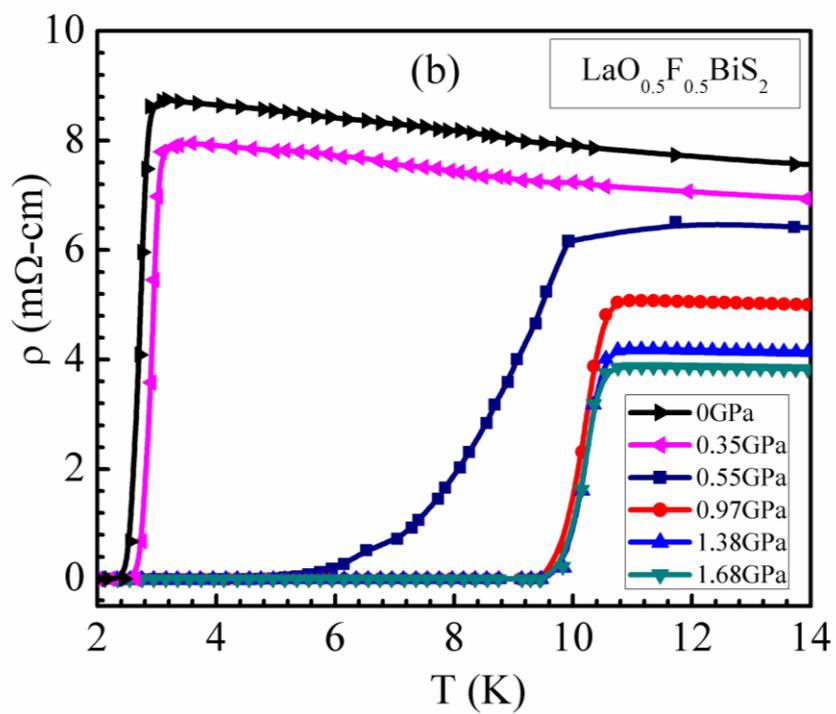

Figure 3

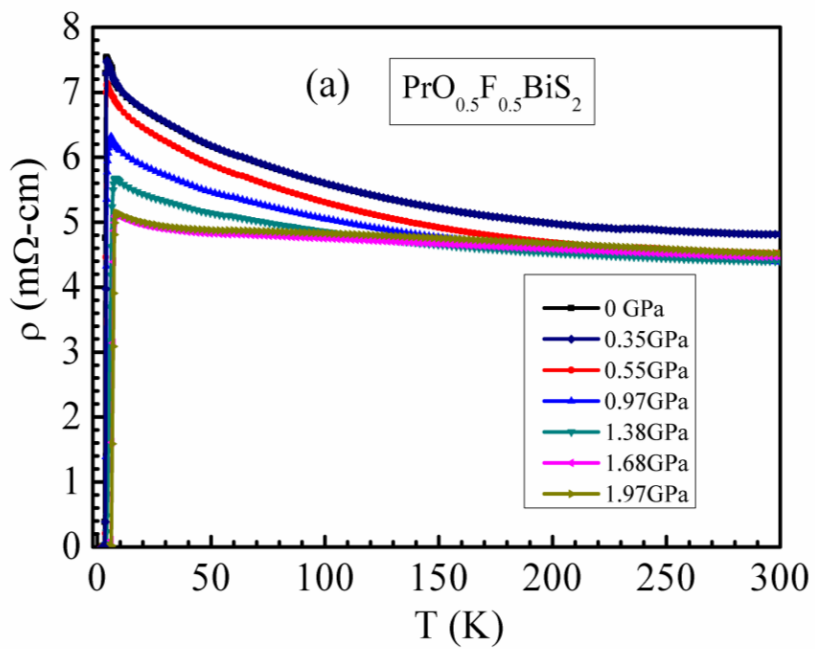

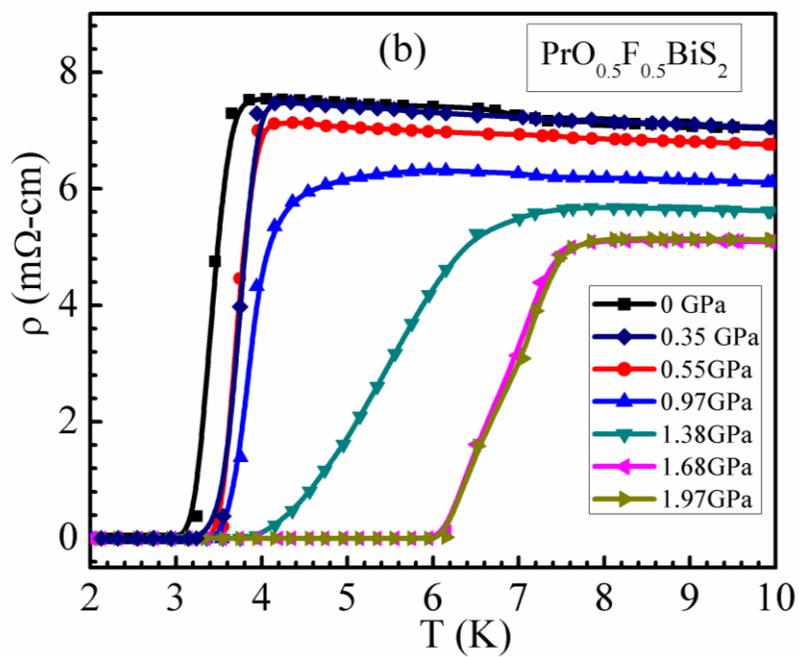

Figure 4

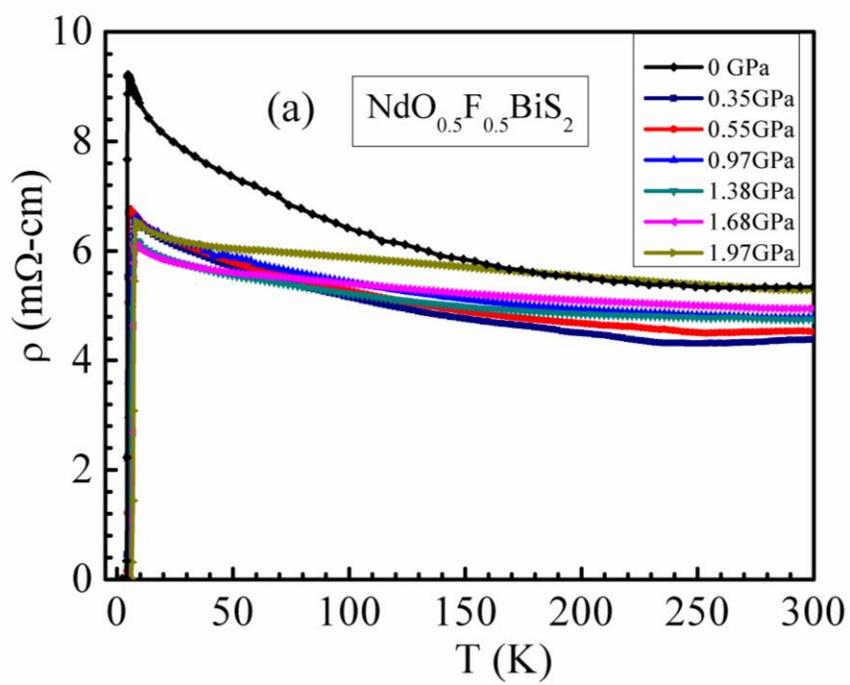

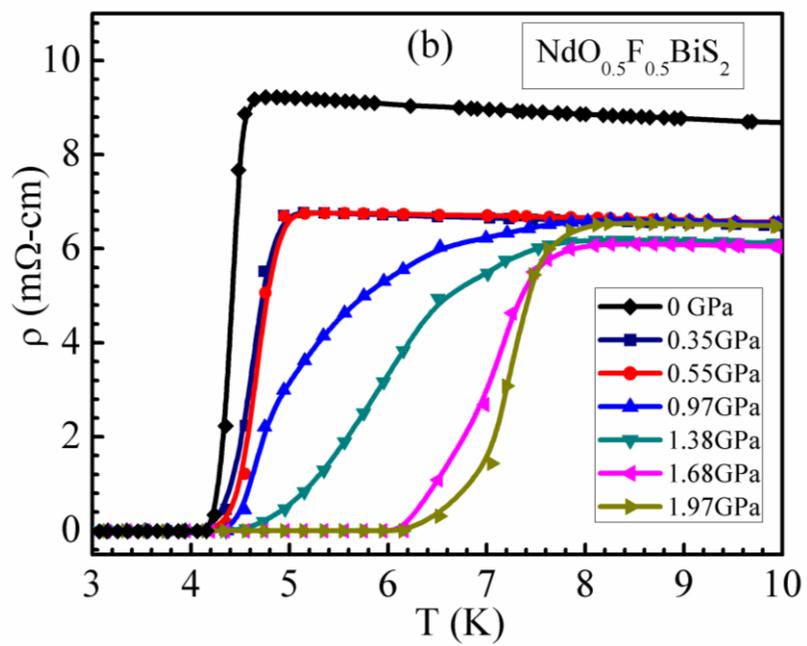

Figure 5

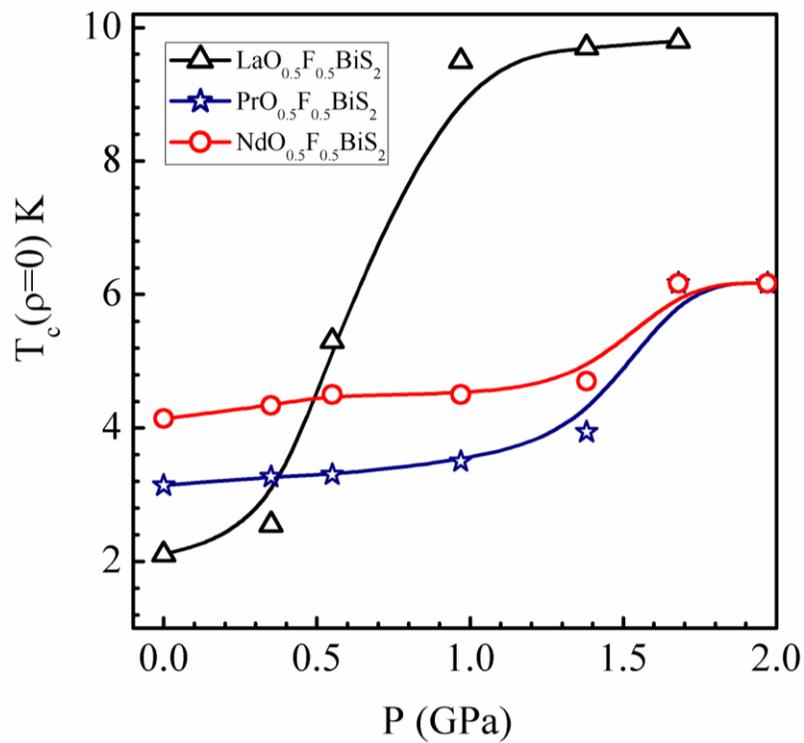

Figure 6

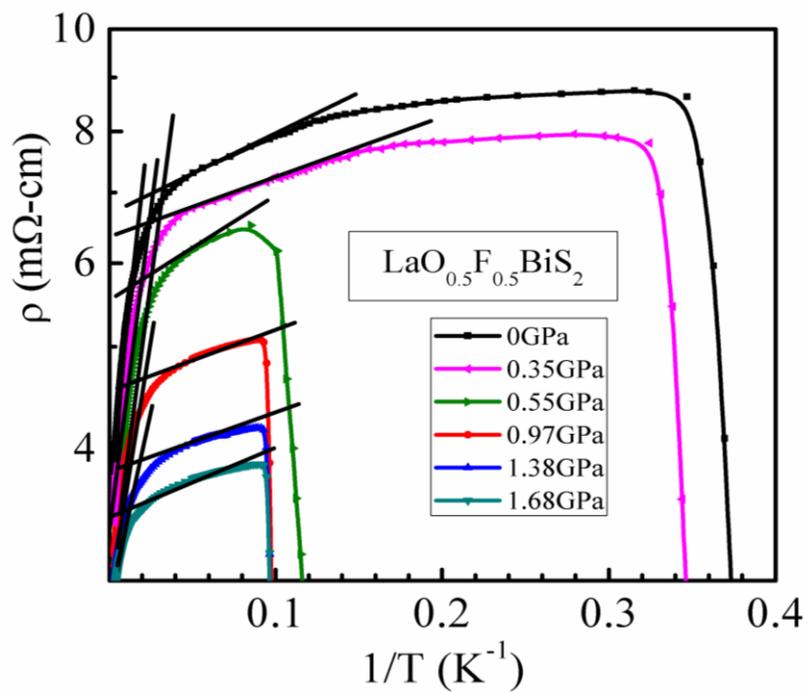

Figure 7

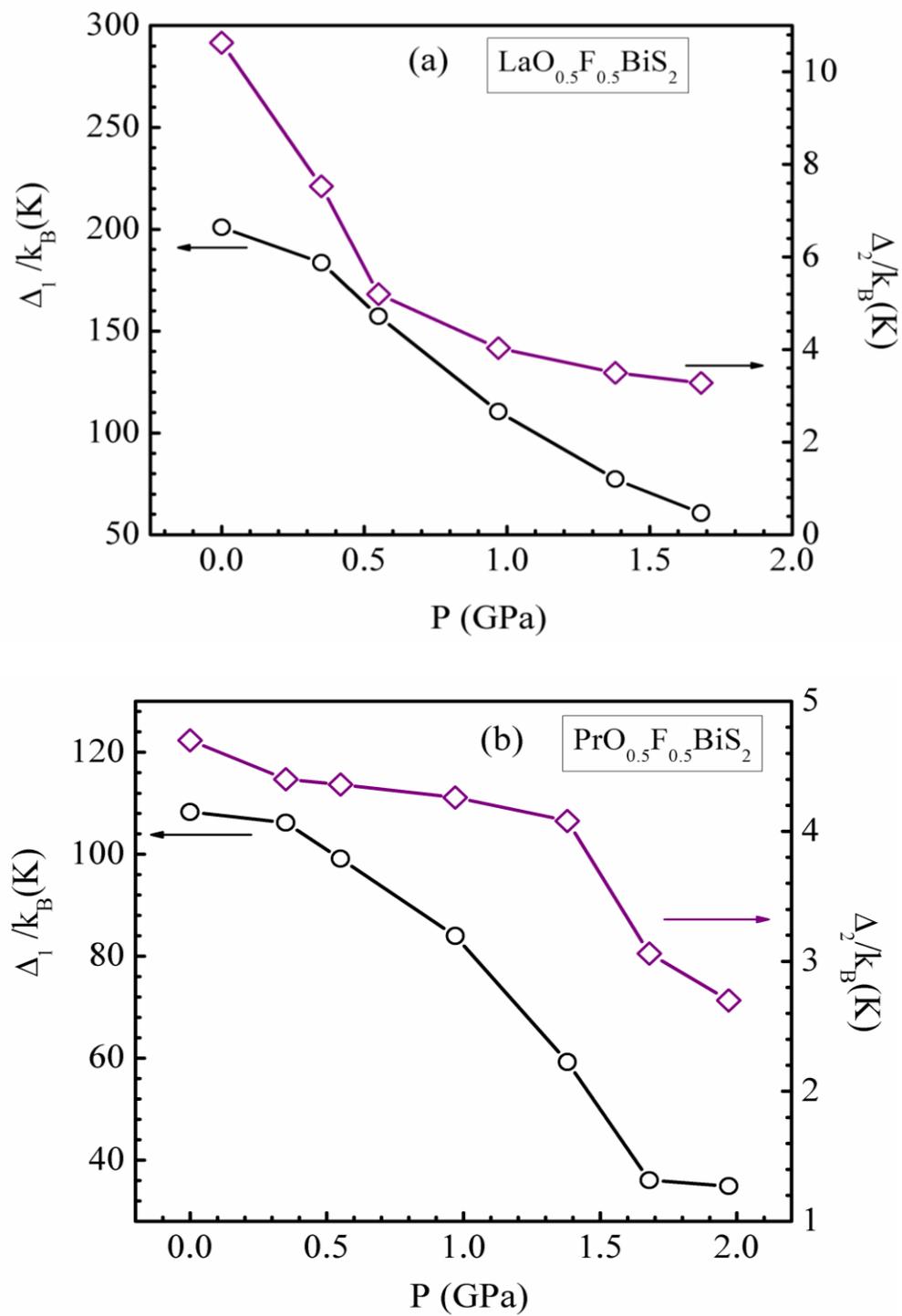

Figure 8

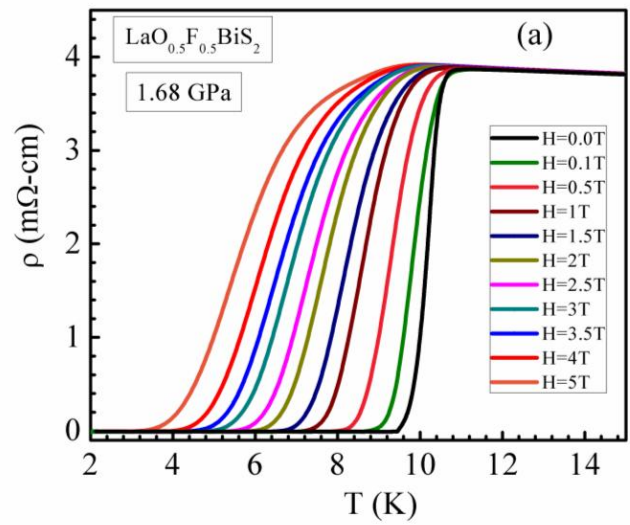

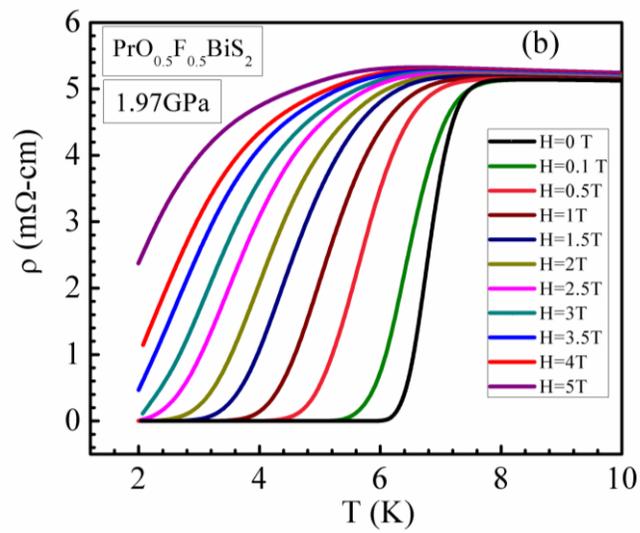

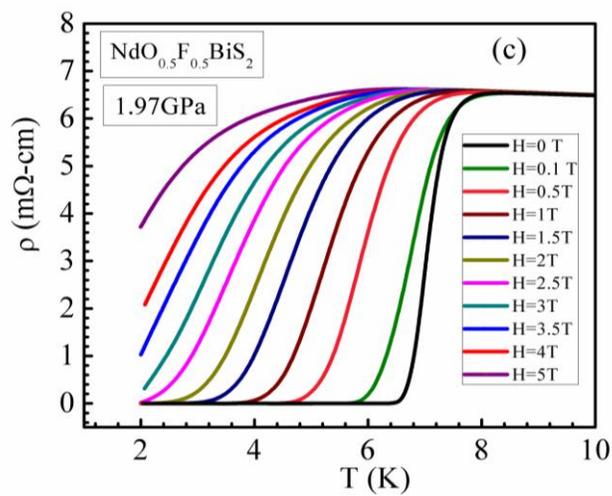

Figure 9

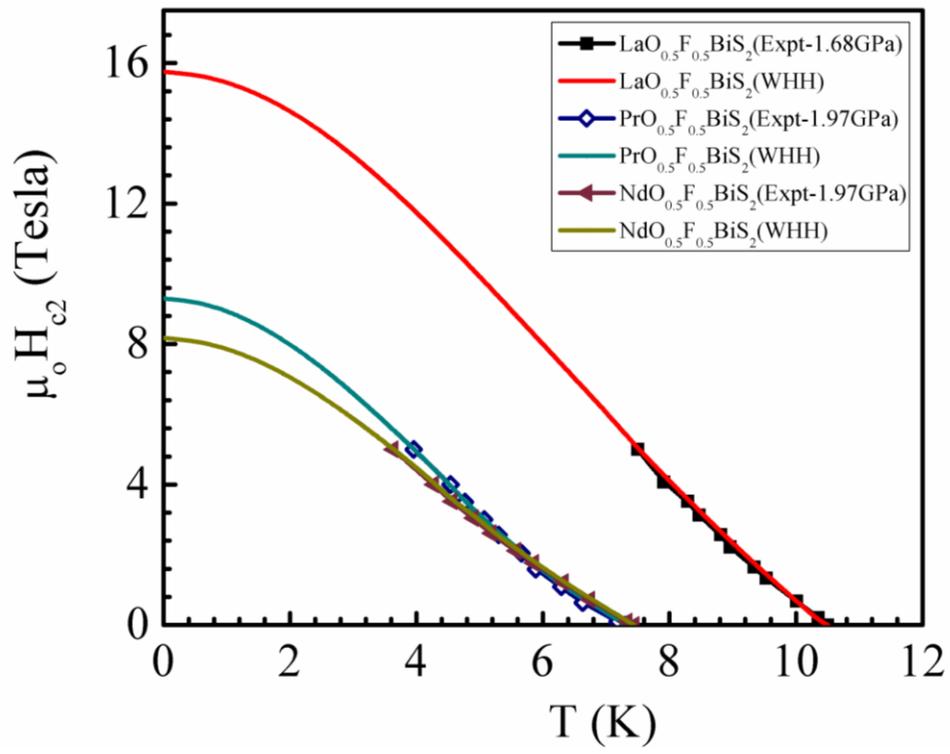